\documentclass[conference]{IEEEtran}
\IEEEoverridecommandlockouts

\usepackage{cite}
\usepackage{amsmath,amssymb,amsfonts}
\usepackage{graphicx}
\usepackage{textcomp}
\usepackage{braket}
\usepackage{tikz}
\usepackage{subcaption}
\usepackage{xcolor}
\usepackage{comment}
\usepackage[ruled]{algorithm2e}
\usepackage{booktabs}

\def\BibTeX{{\rm B\kern-.05em{\sc i\kern-.025em b}\kern-.08em
    T\kern-.1667em\lower.7ex\hbox{E}\kern-.125emX}}
\begin{document}

\title{Decentralized Framework for Teleportation in Quantum Core Interconnects\\
}

\author{
\IEEEauthorblockN{
Rajeswari Suance P S\IEEEauthorrefmark{1},
Ruchika Gupta\IEEEauthorrefmark{2},
Maurizio Palesi\IEEEauthorrefmark{3},
John Jose\IEEEauthorrefmark{1}
}
\IEEEauthorblockA{
\IEEEauthorrefmark{1}Indian Institute of Technology Guwahati, India \quad
\IEEEauthorrefmark{2}Chandigarh University, India \quad
\IEEEauthorrefmark{3}University of Catania, Italy
}
\IEEEauthorblockA{
\textnormal{\{s.rajeshwari, johnjose\}@iitg.ac.in, ruchikae7396@cumail.in, maurizio.palesi@unict.it}
}
}

\maketitle

\begin{abstract}
Multi-core quantum computing architectures offer a promising and scalable solution to the challenges of integrating large number of qubits into existing monolithic chip design. However, the issue of transferring quantum information across the cores remains unresolved. Quantum Teleportation offers a potential approach for efficient qubit transfer, but existing methods primarily rely on centralized interconnection mechanisms for teleportation, which may limit scalability and parallel communication. We proposes a decentralized framework for teleportation in multi-core quantum computing systems, aiming to address these limitations. We introduce two variants of teleportation within the decentralized framework and evaluate their impact on reducing end-to-end communication delay and quantum circuit depth. Our findings demonstrate that the optimized teleportation strategy, termed two-way teleportation, results in a substantial 40\% reduction in end-to-end communication latency for synthetic benchmarks and a 30\% reduction for real benchmark applications, and 24\% decrease in circuit depth compared to the baseline teleportation strategy. These results highlight the significant potential of decentralized teleportation to improve the performance of large-scale quantum systems, offering a scalable and efficient solution for future quantum architectures.

\end{abstract}


\section{Introduction}
Quantum computing is a transformative technology that has the potential to solve the complex problems that classical computers are incapable of solving in deterministic polynomial time. The ability to leverage superposition, interference, and entanglement is expected to give quantum computers a significant advantage in areas such as cryptography, big data,  chemistry optimization, and machine learning \cite{smith2022scaling}.

The current era of Noisy Intermediate Scale Quantum (NISQ) computers, based on monolithic chip design, has witnessed considerable progress in recent years. However, their potential remains limited due to challenges in integrating the sufficient number of qubits required for the efficient execution of practical applications such as Shor's algorithm. It is anticipated that thousands, or even millions, of qubits will be necessary to achieve true quantum supremacy\cite{nisq}. The number of qubits that can be integrated into a monolithic chip is limited by crosstalk, reduced yield, challenges in integrating control circuitry, and per-qubit wiring complexity\cite{alarcon2023scalable}. 

To resolve this, recent advancements have highlighted the potential of interconnecting smaller-sized NISQ devices via short-range communication pathways\cite{niu2023low}. This approach aims to construct a Multi-core Quantum Computing architecture and offers a promising solution to the scalability issue\cite{smith2022scaling}, \cite{alarcon2023scalable}, \cite{palesi2024assessing}. Increasing the isolation between cores reduces cross-talk and impact of correlated errors. By fabricating and assembling smaller units, higher manufacturing yields can be attained as well\cite{wilen2021correlated}. However, communication and coordination among cores constitute a primary challenge in the development of multicore quantum computing architectures\cite{palesi2024assessing}. 

Despite these challenges, multi-core quantum computing architecture has the potential to solve the real-world problems by scaling the number of qubits and facilitating the sharing of quantum information across cores. However, the fragile nature of qubits restricts their transmission as quantum states through quantum channels directly among cores. Instead, a quantum process called teleportation can be employed to enable the transfer of qubits without their physical movement \cite{bennett1993teleporting}. Quantum teleportation is achieved through the combination of classical communication concepts and the use of quantum entanglement to transfer the qubit.

Existing studies mainly emphasizes centralized interconnection mechanism to enable teleportation in the multi-core quantum computing architectures \cite{palesi2024assessing}, \cite{rodrigo2021modelling}, \cite{rodrigo2021double}. In these systems, a central entity manages the generation and distribution of entanglement, which are crucial for teleportation. While the centralized approach simplifies the coordination of entanglement resources, it also introduces challenges related to scalability. In particular, centralization creates bottleneck that restricts parallel communication. As the communication latencies increases, the risk of data loss due to quantum state decoherence becomes more pronounced. To address these limitations of the present model, we propose the following contributions to enhance the interconnection mechanism for teleportation in multi-core quantum computing architectures and minimize the end-to-end communication delay.

\begin{itemize}
    \item We propose a decentralized framework for quantum teleportation in a multi-core quantum computing architecture. 
  \item We introduce two variants of teleportation strategy within proposed framework to reduce both end-to-end communication delay and depth of the quantum circuit.
\end{itemize}

The rest of this paper is organized as follows. Section II covers quantum computing, algorithms, circuit mapping, and communication. Section III discusses the motivation behind the work. Section IV presents the proposed decentralized framework in multi-core architecture and communication protocols based on quantum teleportation. The experimental analysis and their discussion are presented in Section V. Finally, Section VI summarizes the conclusions and future works.

\section{Background}


\subsection{Quantum Computing}
A qubit encodes information in a quantum state with two distinct levels, $\ket{0}$ and $\ket{1}$, and exists in a superposition of both states. This superposition is a key element of quantum computational power. However, qubits are prone to decoherence when exposed to their environment, requiring them to be maintained at cryogenic temperatures.

When measured, a qubit’s state collapses to a classical value, and the no-cloning theorem prevents qubits from being copied, making retransmission impossible\cite{palesi2024assessing}. Another key property is the quantum entanglement which links qubits in such a way that measuring one qubit instantly determines the state of the other, regardless of the distance between them\cite{bennett1993teleporting}. A more comprehensive understanding of quantum computing can be obtained from \cite{national2018quantum}. 

\subsection{Quantum Algorithms and Circuit Mapping}

Quantum algorithms consist of instructions executed on qubits to solve problems. The fundamental unit of a quantum algorithm is the quantum gate, which modifies the quantum states of qubits, similar to classical gates. A quantum algorithm can, therefore, be represented as a sequence of quantum gates, forming a quantum circuit. Computation within this circuit involves one-qubit and two-qubit gates. Gates on different qubits can be applied simultaneously, provided the sequence of operations involving multiple qubits is followed. However, gates acting on the same qubits must be executed sequentially. This leads to the concept of a layer,  which refers to a set of gates that can be executed simultaneously on separate qubits. The total number of layers in a quantum circuit determines its depth. Increased circuit depth results in higher execution times, as it is related to higher communication latency.

For two-qubit gates, there is a strict requirement that both qubits involved must be adjacent. When the input qubits of a two-qubit gate are non-adjacent, they must be moved until they become adjacent. Quantum circuit mapping addresses this by rearranging logical qubits to ensure that, when mapped onto a quantum processing unit, the connectivity constraints do not hinder the operation of two-qubit gates \cite{pastor2024circuit}. However, quantum circuit mapping does not provide a perfect mapping where all the inputs of two-qubit gates are always adjacent. There are instances where the qubits are distributed across multiple cores, requiring an inter-core communication to bring them together for the execution of the two-qubit gate \cite{bandic2023mapping}.

\subsection{Quantum Communication}
Quantum state transmission can be achieved through various techniques. A SWAP gate, for instance, facilitates the exchange of states between adjacent qubits. A qubit can be transferred across cores by chaining multiple SWAP operations. Ion shuttling, specific to ion-trap qubit technology, uses electromagnetic fields to physically move qubits between locations \cite{kaushal2020shuttling}. Another technique, called quantum teleportation, transfers the quantum state to a destination core physically without moving the qubit itself, leveraging classical communication and quantum entanglement \cite{bennett1993teleporting}.

\vspace{-2mm}
\begin{figure}[htbp]
\centerline{\includegraphics[width=90mm, scale=0.75]{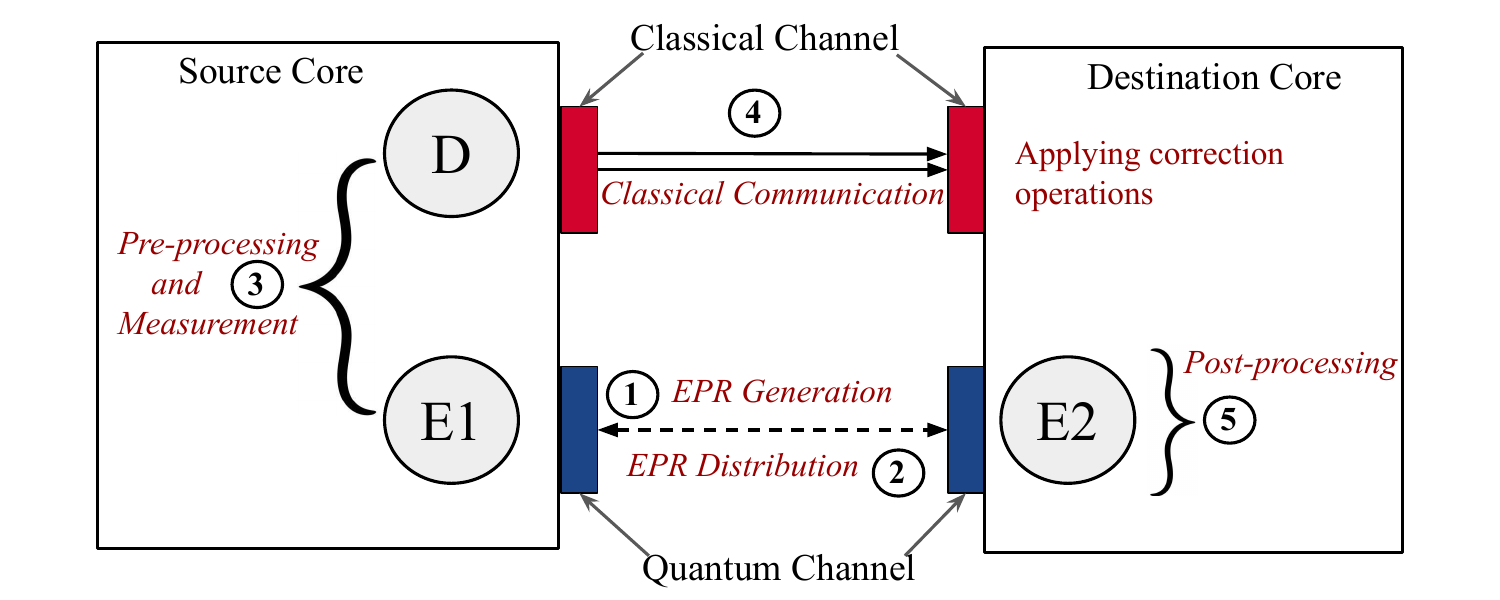}}
\caption{Quantum Teleportation}
\label{teleportation}
\end{figure}
\vspace{-2mm}

Since qubits decohere over time, direct transmission of qubits is not suitable. Instead, quantum teleportation makes use of the fundamental property of qubits called entanglement for communication. The process of quantum teleportation is pictorially represented in Fig. \ref{teleportation}. Once the request for communication is made, two entangled qubits called EPR (Einstein–Podolsky–Rosen) pairs, $E1$ and $E2$, are generated and distributed through the quantum channel to the source and destination as shown in step 1 and 2 of the figure. Following that, in step 3, a pre-processing and measurement are performed on one of the EPR pairs, $E1$ and the required qubit $D$ in the source core. This step results in two classical bits of information, which are transmitted through the classical channel to the destination core in step 4. Finally, in step 5, post-processing is carried out at the destination core using the two received classical bits and the other EPR pair $E2$, ultimately reconstructing the desired qubit at the destination. This approach ensures that no-cloning theorem is respected. 

Quantum communication is highly sensitive to latency. When the communication latency exceeds the qubit's decoherence time, the computation's error rate increases significantly, leading to incorrect results.

\section{Motivation}

Three distinct approaches can be employed for entanglement generation in the teleportation process: (i) utilizing a shared centralized EPR pair generator across all nodes; (ii) a decentralized approach integrating a Bell State Measurement device at each core-to-core connection; and (iii) generating entanglement at the source, where an entangled pair is produced at the transmitter node, and one of the photons is sent to establish entanglement with a remote node \cite{cacciapuoti2020entanglement}. 

\begin{figure}[htbp]
\centerline{\includegraphics[width=60mm, scale=0.75]{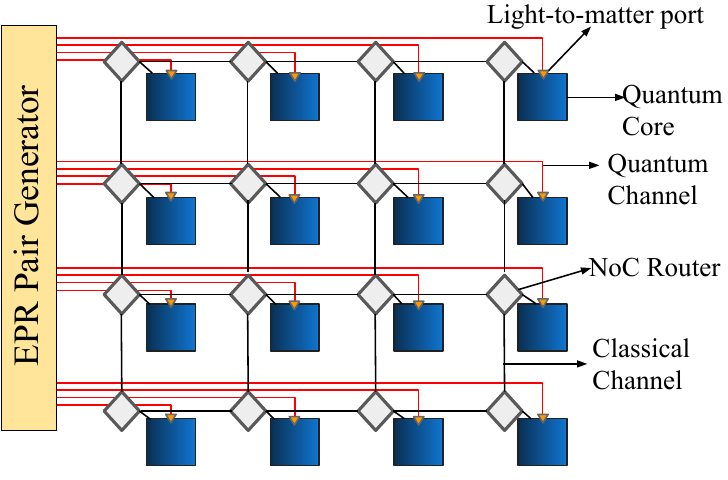}}
\caption{Centralized Interconnection Framework}
\label{centralized}
\end{figure}

The existing works \cite{rodrigo2021modelling, palesi2024assessing, rodrigo2021double} predominantly adopt the centralized interconnection architecture to facilitate teleportation. A key bottleneck in this approach is the centralized EPR pair generator and the point-to-point connection between the EPR pair generator and the cores through light-to-matter ports as shown in Fig.~ \ref{centralized}. As the number of cores increases, these architectures face the scalability challenges due to the limitation of point-to-point connections to the EPR pair generator.

EPR pair is required to communicate between two cores in the quantum teleportation. If two of the cores want to communicate, then they must send a request to the EPR pair generator. The centralized EPR generator produces EPR pair and distribute through the point-to-point connection by a light-to-matter port. Both the centralized entity and the limited light-to-matter ports reduce the potential for parallel communication. In contrast, we propose an efficient framework based on a decentralized method for generating the EPR pairs.

The simplest method for transferring qubits is swapping. However, the swapping is limited to smaller quantum computing architecture and increases the gate count, which directly affects the depth of the circuit, and consequently communication delay \cite{hillmich2021exploiting}. In turn, quantum teleportation can be used for larger architectures \cite{oskin2003building}. Teleportation utilizes quantum phenomena, such as entanglement, to a level that classical methods, such as swapping, cannot replicate. 

\section{Decentralized Framework for Teleportation}

\begin{figure}[htbp]
\centerline{\includegraphics[width=60mm, scale=0.75]{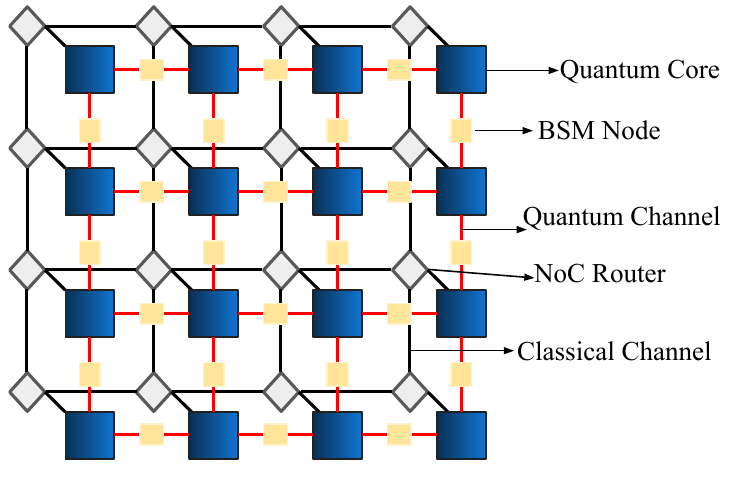}}
\caption{Decentralized Interconnection Framework}
\label{decentralized}
\end{figure}

The proposed decentralized framework for multi-core quantum computing architecture is depicted in Fig.~\ref{decentralized}. It comprises of 2D mesh of quantum cores in $x$- (horizontal)  and $y$- (vertical) dimensions. These quantum cores are interconnected by Bell State Measurement (BSM) nodes and a dual quantum and classical Network-on-Chip (NoC). Within each core, $N$ computation qubits are dedicated to the computational tasks, while $M$ communication qubits are assigned to handle inter-core communication requests. We assume an all-to-all connectivity among the qubits in the core. Each core is connected to its four neighbor cores in south, north, east, and west direction, except for the corner cores, which are only connected to just two neighbors. A BSM node is inserted in between every pair of adjacent cores. Entanglement required for the teleportation of qubits in an inter-core communication is facilitated by the BSM nodes, utilizing both quantum and classical NoC.

In this architecture, the quantum NoC connects the quantum cores to their BSM nodes through optical channel. Quantum NoC is responsible for transmitting photons emitted from the qubits to generate the entanglement between the cores. Meanwhile, the classical NoC connects the quantum cores through routers, enabling the exchange of classical bits using traditional communication methods. Specifically, it generates control signals to initiate inter-core communication, determines the optimal routing path for successful qubit transfer, and ensure overall synchronization of quantum operations.

\begin{figure}[htbp]
\centerline{\includegraphics[width=65mm, scale=0.75]{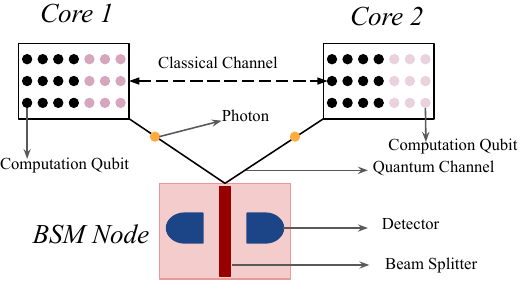}}
\caption{Entanglement Generation Between Two Cores}
\label{barret-kok}
\end{figure}

Fig.~\ref{barret-kok} illustrates entanglement generation based on the Barret-Kok generation principle \cite{barret_kok}. During this process, a photon is emitted from the communication qubit of each cores and directed towards the intermediate BSM node through quantum channel. The BSM node detects the incoming photons and carries out a Bell state measurement. If the measurement is successful, the communication qubits in the cores become entangled enabling quantum communication.

\vspace{-3mm}
\vspace{-2mm}




Now, the role of compiler is to map quantum circuits onto the underlying architecture optimally, while adhering to the given connectivity constraints between qubits. Despite the fact that the compiler attempts to find the most efficient mapping, it is not always possible to place the inputs of a two-qubit gate within the same core. In such cases, inter-core communication is necessary to facilitate the interaction between qubits that are distributed across different cores. We propose two distinct methods of teleportation, namely \emph{hop-by-hop teleportation} and \emph{two-way teleportation}, to facilitate this communication task.

\subsection{Hop-by-Hop Teleportation}

This is the baseline communication infrastructure for the decentralized interconnection framework. The hop-by-hop teleportation method follows the deterministic XY routing algorithm to determine the path between source core and destination core in an inter-core communication. 

\begin{figure}[htbp]
\centerline{\includegraphics[width=45mm, scale=0.75]{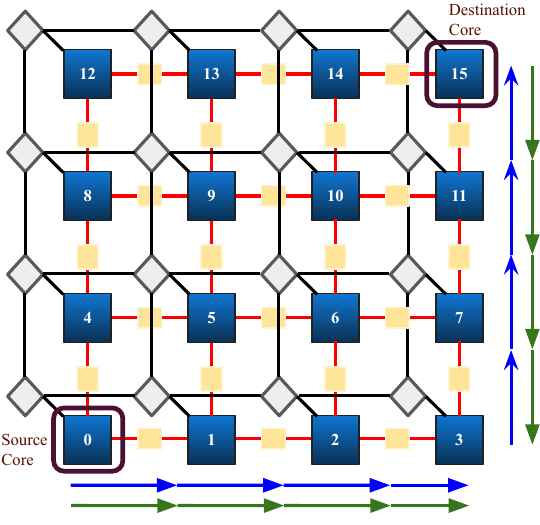}}
\caption{Hop-by-Hop (blue arrows) and Two-Way (green arrows) Teleportation}
\label{example_hh_twt}
\end{figure}

When the source core and destination core are one hop apart, the entanglement is generated using the intermediate BSM nodes. The qubit required for the two-qubit gate in the source core is then moved to destination core using the quantum teleportation. If there is more than one hop between the source and destination cores---for example, between core 0 and core 15 in Fig.~\ref{example_hh_twt}---the source core transfers the qubit to the next core along the path determined by the XY routing algorithm, as indicated by the blue arrows in the figure.
This is done using the decentralized entanglement generation protocol and quantum teleportation. This process is repeated until the qubit reaches the destination core, where the two-qubit gate can be executed successfully with the input qubits adjacent to each other. However, this can lead to higher communication delay since the movement of the qubit is one-way.

\vspace{-3mm}
\vspace{-4mm}

\subsection{Two Way Teleportation}
In hop-by-hop teleportation, only one qubit moves at a time toward the destination core. This process is time-consuming, especially when executing a two-qubit gate at the destination core. Additionally, it can lead to potential data loss due to quantum state decoherence. Two-way teleportation is an optimization of hop-by-hop teleportation designed to reduce the end-to-end communication delay. 

In two-way teleportation, whenever an inter-core communication request is received, both the source and destination cores begin generating entanglement with their neighboring cores. Through quantum teleportation, the qubits needed for the two-qubit gate at both the source and destination cores move one step closer, as shown by the green arrows in Fig.~\ref{example_hh_twt}.
This effectively reduces the distance between the required qubits. The process continues iteratively, with qubits moving from both sides until they become adjacent at an intermediate core (e.g., core 3 in Fig.~\ref{example_hh_twt}).
Unlike hop-by-hop teleportation, where the two-qubit gate is executed at the destination core, in two-way teleportation, the gate execution occurs at the intermediate core closer to the destination.
\begin{algorithm}
    \caption{Two Way Teleportation}
\label{alg:two_way}
\KwIn{Direction of destination core w.r.t source core $dir\_dest$}
\KwOut{Final traversal direction from source core and destination core $tele\_src\_dir,  tele\_dest\_dir$}
\SetKwFunction{FMain}{Two Way Teleport}
\SetKwProg{Fn}{Function}{:}{}
   \While{(qubits reach middle\_core)}{
       \If{$dir\_dest$ is in [East, West]}{
            $tele\_src\_dir$ = $tele\_dest\_dir$ = X
       }
       \ElseIf{$dir\_dest$ is in [North, South]}{
            $tel\_src\_dir$ = $tel\_dest\_dir$ = Y
       }
       \ElseIf{$dir\_dest$ is in [East-North, East-South, West-North, West-South]}{
            $tele\_src\_dir$ = X

            $tele\_dest\_dir$ = Y
       }
}
\end{algorithm}

The routing algorithm for two-way teleportation is described in Algorithm~\ref{alg:two_way}. The direction of teleportation is determined dynamically using the XY routing algorithm. If the source core and destination cores lie along the same direction, teleportation proceeds along that dimension. When the destination core is diagonally placed (East-North, East-South, West-North, West-South), the source core moves accordingly in the X direction, while the destination core moves in the Y direction, as determined by the algorithm. 
\vspace{-2mm}
\begin{figure*}[h!]
    \centering
    \begin{subfigure}[t]{0.30\textwidth}
        \centering
        \includegraphics[width=60mm, scale=0.75]{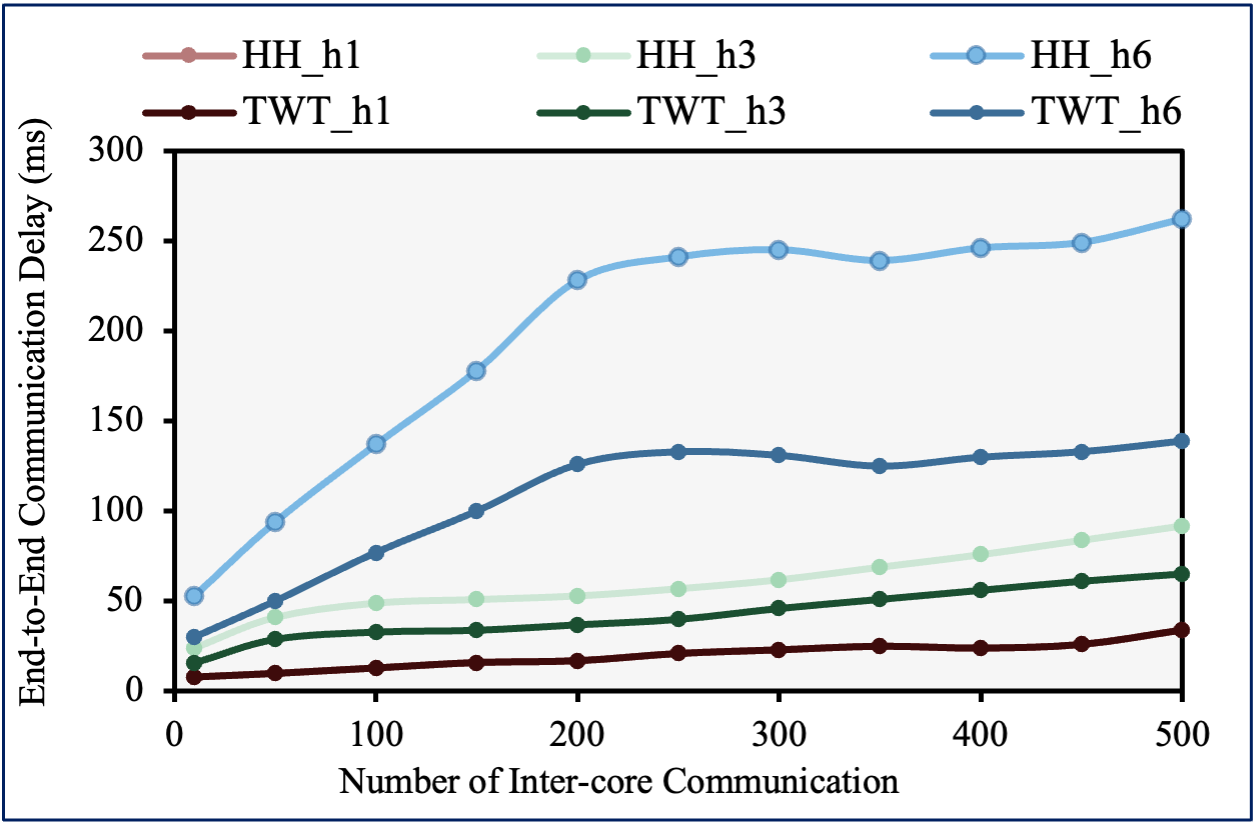}
        \caption{Fixed $C_r$  with Depth 5}
        \label{theoretical_study}
    \end{subfigure}
    \hspace{0.03\textwidth} 
    \begin{subfigure}[t]{0.30\textwidth}
        \centering
        \includegraphics[width=60mm, scale=0.75]{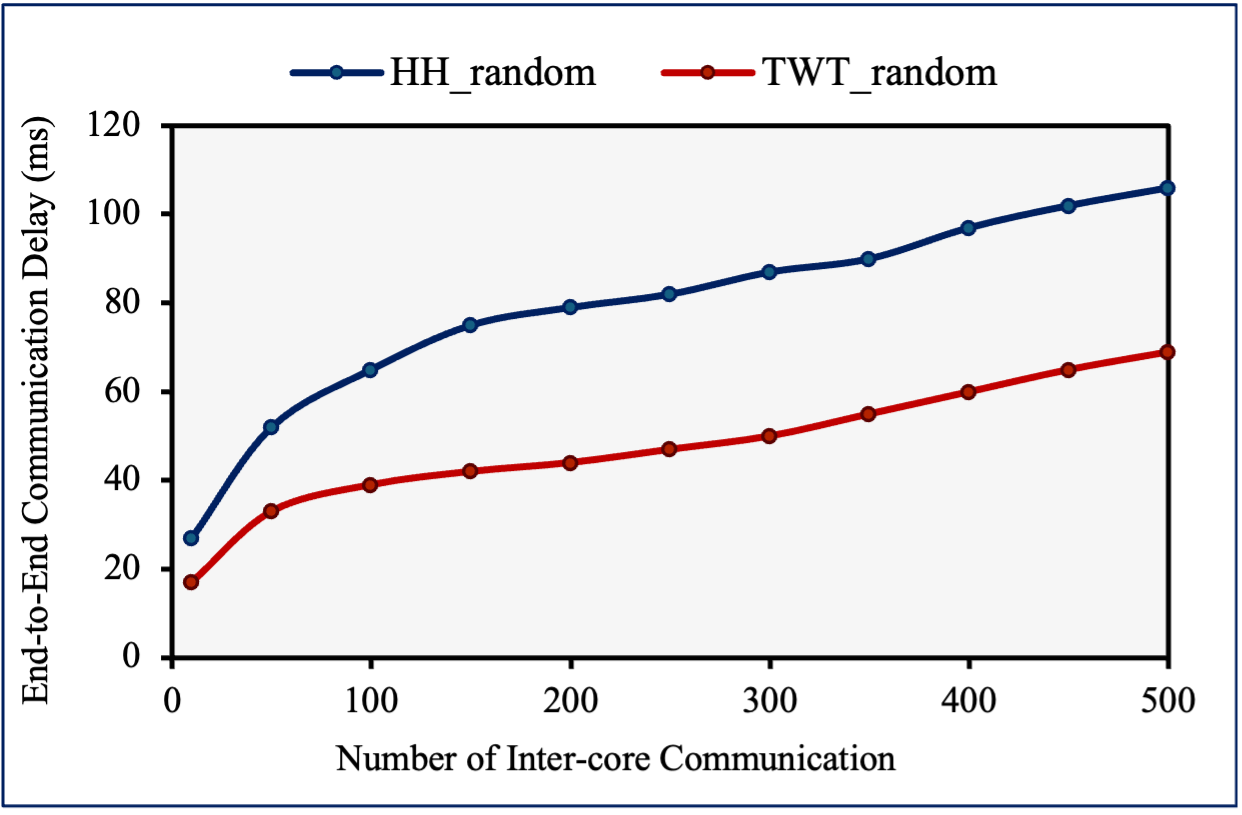}
        \caption{Random $C_r$ with Depth 5}
        \label{d5_random}
    \end{subfigure}
    \hspace{0.03\textwidth} 
    \begin{subfigure}[t]{0.30\textwidth}
        \centering
        \includegraphics[width=60mm, scale=0.75]{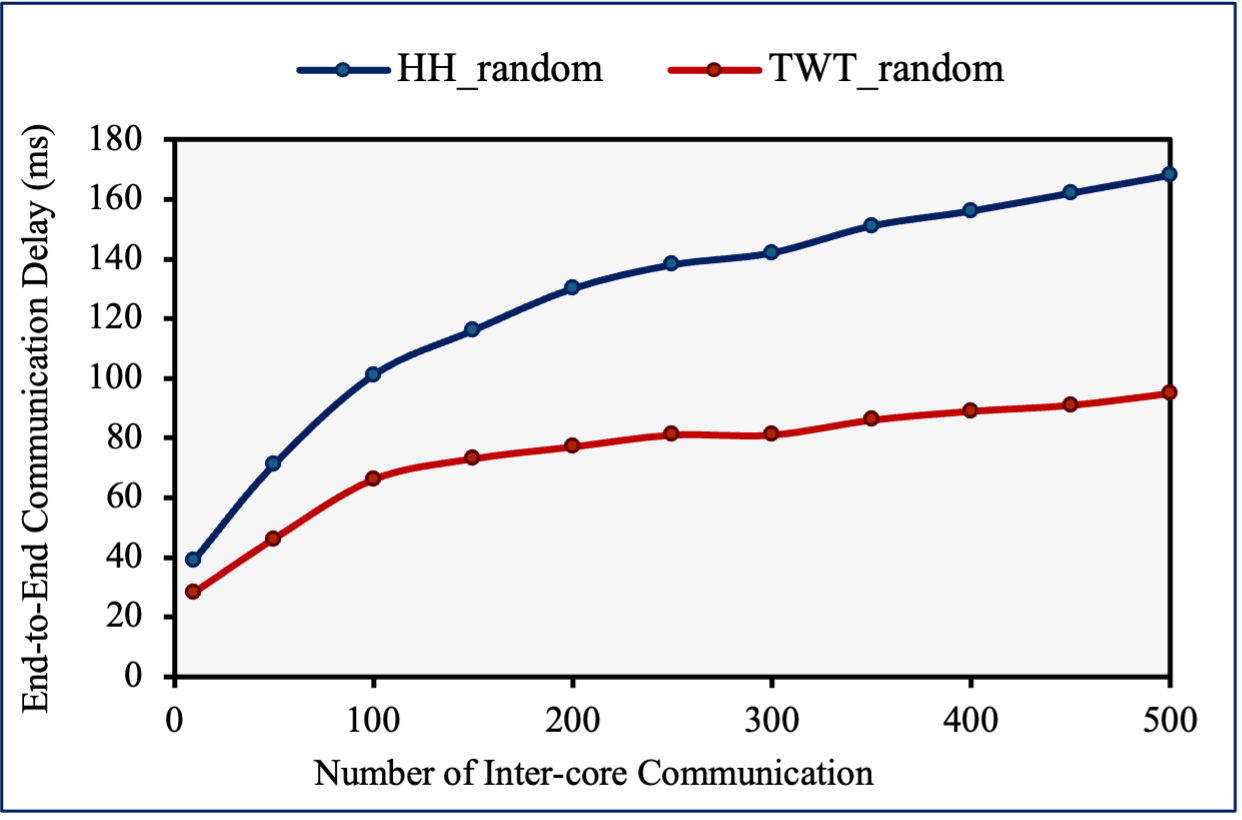}
        \caption{Random $C_r$ with Depth 10}
        \label{d10_random}
    \end{subfigure}
    \caption{End-to-end Communication Delay  versus the Number of Inter-core Communications for Synthetic Benchmarks}
\end{figure*}

\section{Evaluation}
\subsection{Experimental Setup}
We simulate quantum circuits on the decentralized multi-core quantum architecture to analyze the impact of inter-core communication. To set up the simulation, we modify SeQUeNCe, an open-source quantum network simulator \cite{sequence}. Specifically, we develop a quantum teleportation protocol to facilitate qubit transfer efficiently between cores using classical communication and entanglement. We consider 4x4 2D mesh architecture. We assume there is an all-to-all connectivity among qubits inside the core. We evaluate the quantum circuit mapping as a direct 1:1 mapping of logical qubits to physical qubits on the device, specifically without considering any compiler optimization of the mapped circuit. 

\textbf{\textit{Metrics}}: When the input of two-qubits gates are distributed among many cores, there is a need for inter-core communication. We experiment with scaling the number of inter-core communication requests in a circuit to assess the end-to-end communication delay required using the proposed methods in the decentralized framework. Additionally, we consider the depth of the compiled circuit as a metric since it indicates the latency of program execution. The circuit depth also impacts the decoherence time of the qubits.

\textbf{\textit{Benchmarks}}:  
We implement random circuits as a synthetic benchmark to effectively control the parameter of inter-core communication requests. Only two-qubit gates are considered, where the inputs are located on different cores. We conduct experiments on the random circuit with depths of 5 and 10. Additionally, subroutines, which are building blocks of real application such as the Quantum Fourier Transform (QFT)\cite{ruiz2017quantum}, the Multi-Target Gate, Cuccaro’s Ripple-Carry Adder\cite{cuccaro2004new} are used as real benchmarks. We have also included Quantum Volume, which involves high gate density circuits, under the real benchmarks. These benchmarks are generated using Qiskit circuit implementation.

\subsection{Experimental Analysis}
We define a variable called Connectivity Radius $C\textsubscript{r}$, where the $r$ denotes the number of hops from the source core to destination core in an inter-core communication.  We conduct a theoretical study on the synthetic benchmark, considering three different scenarios, as shown in Fig.~\ref{theoretical_study}, which illustrates the end-to-end communication delay versus the number of inter-core communications.
We compare the hop-by-hop and two-way cases, referred to as HH\_h$n$ and TWT\_h$n$, where $n$ represents the selected $C\textsubscript{r}$ value. (i). When $C\textsubscript{r}=1$, one of best possible quantum circuit mapping from the compiler is achieved, i.e., the source and destination are neighbors. In this case, both hop-by-hop and two-way teleportation behave similarly; (ii). When $C\textsubscript{r}=3$, the average distance between the source and destination core is 3. Here, two way teleportation begins to outperform hop-by-hop teleportation, reducing the end-to-end communication delay by 29\%; (iii). When $C\textsubscript{r}=6$, it represents the maximum distance possible between the source and destination core in the considered 4x4 mesh architecture, which corresponds to one of worst case scenario of the quantum circuit mapping. In this case, two-way teleportation reduces the end-to-end communication latency by an average of 84\% compared to hop-by-hop teleportation.
\vspace{-3mm}
\begin{figure}[htbp]
\centerline{\includegraphics[width=60mm, scale=0.75]{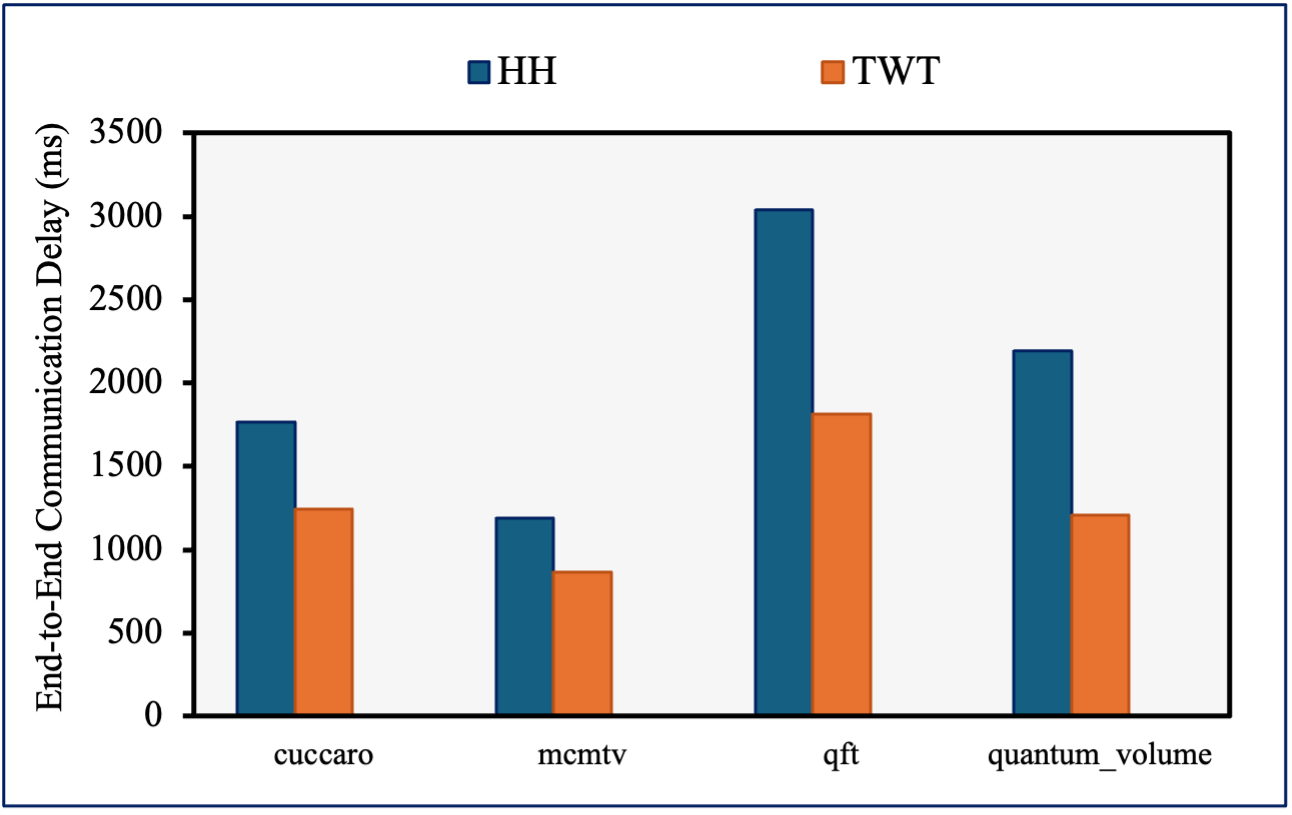}}
\caption{End-to-end Communication Delay for Real Benchmarks}
\label{real_comm}
\end{figure}
\vspace{-2mm}
For a practical scenario, we execute the synthetic benchmark with randomly assigned $C\textsubscript{r}$ for each inter-core communication requests. It was observed that the two way teleportation reduces the end-to-end communication delay by an average of 40\% outperforming hop-by-hop teleportation for both the circuit depths of 5 and 10, as shown in Fig. ~\ref{d5_random} and \ref{d10_random}.

When we run experiments on real benchmarks, as shown in the Fig.~\ref{real_comm}, two-way teleportation (TWT) consistently reduces the end-to-end delay compared to hop-by-hop teleportation (HH) across all benchmarks. For Cuccaro's Adder and MCMTV, a 30\%, while QFT and Quantum Volume show an average reduction of 42\% in end-to-end communication delay. The greater reduction in communication delay in QFT is due to the fact that two-qubit inter-connectivity involves connecting each qubit to all other qubits, except for itself. Whereas in Cuccaro's adder, the two-qubit inter-connectivity is limited to nearest neighbors \cite{rached2023characterizing}. 

\begin{figure}[htbp]
\centerline{\includegraphics[width=60mm, scale=0.75]{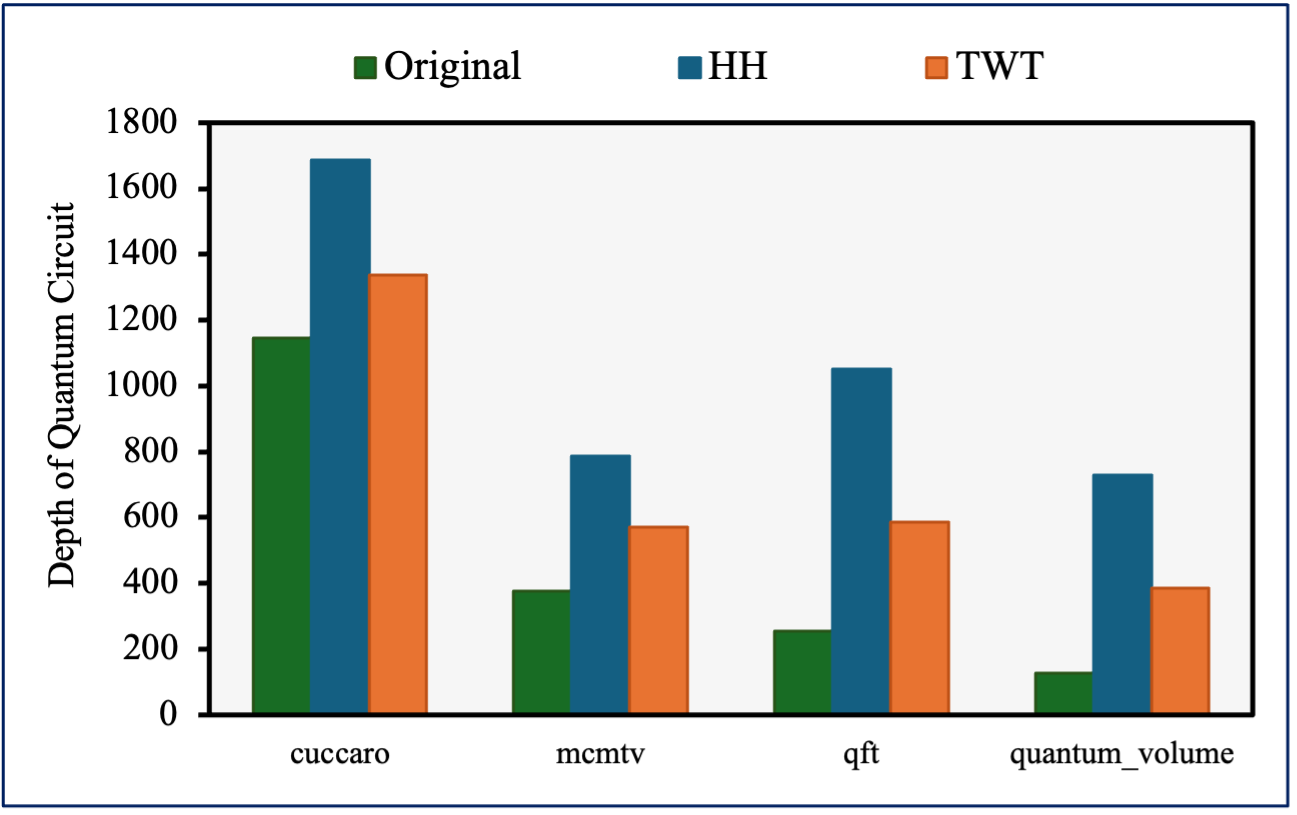}}
\caption{Quantum Circuit Depth Analysis}
\label{real_depth}
\end{figure}

Depth of the quantum circuit is reduced to 24\% in case of Cuccaro's Adder and MCMTV while 45\% reduction can be observed for QFT and Quantum Volume when the two-way teleportation (TWT) is employed than hop-by-hop (HH) communication as shown in Fig.~\ref{real_depth}. Compared to the original depth of the circuit, two-way teleportation does not increase the depth in the same way as hop-by-hop communication.
\vspace{-3mm}
\vspace{-3mm}
\subsection{Discussion}
The experimental results shows that two-way teleportation reduces the end-to-end communication delay compared to the hop-by-hop teleportation. This is due to the fact that, both qubits from the source and destination cores move towards the middle core simultaneously, where the two-qubit gate is executed. In contrast, in the hop-by-hop approach, only the qubit from the source core gradually moves towards the destination core at any given time, delaying the execution of two-qubit gate. This sequential movement significantly increases the overall communication latency, and consequently, the execution time of the quantum algorithms, thereby affecting the decoherence time of the qubits.

The depth of the quantum circuit is also reduced in two-way teleportation compared to hop-by-hop teleportation. All gates at depth $d$ must be fully executed before the gates at depth $d+1$ can begin. In the case of inter-core communication, if connectivity radius $C\textsubscript{r} > 2$, each time a qubit is moved closer to the destination core, the circuit depth increases by one due to the constraint of respecting the execution order. In a hop-by-hop teleportation, the depth increases linearly, as the distance to be covered is larger and two-qubit gate execution happens at the destination core. However, in two-way teleportation, both the source core and destination core transfer the qubit simultaneously, which results in depth increase less than that of hop-by-hop communication. 

The main limitation of the two-way teleportation is the potential for congestion in the middle cores. These cores may experience a higher density of qubits, resulting in a potential imbalance on number of qubits they can handle. This issue arises primarily due to the choice of routing algorithm, as we have opted for the deterministic XY routing algorithm. 

Quantum swapping and quantum purification, commonly used in the Quantum Internet \cite {cacciapuoti2020entanglement} are applicable to the proposed architecture. However, the limited availability of on-chip resources restricts their implementation.

\section{Conclusion and Future Works}
The challenges posed by centralized interconnection mechanisms in multi-core quantum computing architectures have been addressed by proposing a decentralized framework for teleportation. We introduced two teleportation variants on the decentralized framework: a baseline hop-by-hop and an optimized two-way teleportation. The two-way teleportation strategy reduced the end-to-end communication delay by 40\% for synthetic benchmarks, 30\% for real benchmarks, circuit depth by 24\% compared to hop-by-hop teleportation. These results highlight the effectiveness of decentralized teleportation in enhancing scalability and performance. 

However, congestion issue may arise due to the deterministic XY routing algorithm used for two-way teleportation. Future work will address this limitation by developing an adaptive routing algorithm to further optimize performance. This will enable more efficient and scalable quantum systems, advancing the potential of decentralized teleportation in large-scale multi-core quantum computing architectures. 

\section*{Acknowledgment}

Authors gratefully acknowledge funding from the European Commission through HORIZON-EIC-2022-PATHFINDEROPEN01-101099697 (QUADRATURE) and University of Catania, Fondi di Ateneo project PIACERI 2024-2026 QUIET.

\bibliographystyle{IEEEtran}
\bibliography{reference}
\end{document}